\documentclass[12pt,preprint]{aastex}

\shorttitle{GRB 060614: a chance superposition?}
\shortauthors{Cobb et al.}

\begin{document}

\title {Could GRB 060614 and its presumed host galaxy be a chance superposition?}

\author{B.~E. Cobb\altaffilmark{1}, C.~D. Bailyn\altaffilmark{1}, P.~G. van Dokkum\altaffilmark{1}, and
P. Natarajan\altaffilmark{1}}
\email{cobb@astro.yale.edu}

\altaffiltext{1}{Department of Astronomy, Yale University, P.O. Box 208101, New Haven, CT 06520}

\begin{abstract}
 The lack of an observed supernova associated with GRB 060614 appears to
 require a new paradigm for the formation of (a subset of) long-duration GRBs. 
 This requirement is based on the presumed low
 redshift of the burst, which was inferred from the spatial
 coincidence of the afterglow with a $z=0.125$ galaxy. We explore
 the possibility that this low-redshift galaxy is a chance
 superposition along the line of sight to GRB 060614. We examine the galaxy
 distribution of the field of GRB 060614 and find that the probability of a chance association
 with a galaxy at least as bright as the putative host is
 only $\sim 0.5 - 1.9$\%. However, for the current ensemble of
 $\approx 180$ \textit{Swift} GRBs it is likely that several such coincidences have occurred, and
 given the ``non-standard" nature of GRB 060614 it is not implausible
 that this is one such occurrence.
 Thus the conclusion that GRB 060614 requires a revision
 to the formation paradigm for long-duration GRBs should
 be approached with caution.

\end{abstract}


\section{Introduction}
The evidence that long-duration GRBs are associated with supernovae (SNe) is very strong.  
From observed low-luminosity GRBs with SNe that dominate the bursts' optical emission (e.g. GRB 980425/SN 1998bw 
and 060218/SN 2006aj; \citealt{Galama+98,Cobb+06,Ferrero+06,Mirabal+06,Modjaz+06,Sollerman+06}) 
to high-luminosity GRBs with SNe embedded in the bursts' optical afterglow 
(e.g. GRB 030329/SN 2003dh; \citealt{Hjorth+03,Stanek+03,Bloom+04}) has emerged a paradigm in which long-duration GRBs are produced 
by the core collapse supernovae of  massive stars. 

The study of this association is difficult at high redshifts, 
due to the intrinsic faintness of the SNe when compared with the luminosities produced at all wavelengths 
by ultra-relativistic GRB jets.  Any possible low-redshift GRB is, therefore, of great importance 
and is scrutinized by many observers for photometric or spectroscopic evidence of a SN.   
Such was the case for GRB 060614 \citep{Parsons+06}, a burst that was typical in most respects and was 
followed by a relatively bright and long-lasting optical afterglow \citep{Holland06,Brown+06}.    

Initial observations of GRB 060614 detected afterglow light in all \textit{Swift} UVOT bands, 
indicating that the GRB had a moderate redshift of $z<1.7$, due to the absence of 
absorption from the Lyman-$\alpha$ forest \citep{Holland06,SX06}.  From the spectral parameters of the GRB, 
\cite{PA06} calculated the pseudo-redshift of GRB 060614 to 
be $pz= 1.45\pm0.85$, consistent with the earlier estimate \citep[see also][]{SX06}.  
Spectroscopic observations of the afterglow of GRB 060614 taken less than 
two days post-burst did not yield a spectroscopic redshift 
\citep{Fugazza+06a}, as no absorption or emission features were detected on top of the 
power-law spectrum of the afterglow.    

More spectra were obtained at the afterglow position of GRB 060614 as the afterglow continued to fade away.  
In a spectrum taken $\sim5$ days post-burst \citep{Price+06}, a single strong emission line was 
detected and interpreted to be due to H$\alpha$ from the GRB host galaxy at $z=0.13$.  
The redshift of this potential host galaxy was confirmed to be $z=0.125$ by \cite{Fugazza+06b}.  
With the possibility of this burst being at very low redshift, several observers continued careful 
follow-up campaigns \citep[e.g.][]{BH06,CB06,Fynbo+06a}, anticipating the brightening expected from a SN at that redshift. 
However, in time, all decaying afterglow light curves leveled off to a constant magnitude, 
interpreted to be the magnitude of the host galaxy, and no SN brightening was observed \citep{CB06,DellaValle+06,Fynbo+06b,Gal-Yam+06}.

The lack of an observed SN from this seemingly low-redshift GRB can be explained in several ways:
\begin{enumerate}
\item A SN did occur but was not detected due to high line-of-sight extinction 
toward the GRB, most likely due to dust in the GRB's host galaxy.  However 
there are no indications of the presence of a supernova in IR observations
\citep{CB06} and also no 
indication of the presence of this reddening in the UV/optical/IR 
observations of the afterglow \citep{Holland06,DellaValle+06,Fynbo+06b,Gal-Yam+06}.
\item A SN did occur but was not detected because the SN was underluminous by more 
than 5 magnitudes.  This would be unexpected, however, as all previous 
GRB-related SNe have luminosities that cluster within a magnitude of the peak
brightness of SN 1998bw \citep{ZKH04,Cobb+06,Ferrero+06}. While local non-GRB Type Ibc SNe
do span a larger range of $\pm\sim2$ magnitudes from the peak of SN 1998bw \citep{Richardson+06}, even
the dimmest known Ibc SN should have been detectable in the absence of strong reddening.
\item No stellar core collapse occurred, or the core collapse did not result in a typical GRB ``hypernova".
These possibilities have now been suggested by several authors 
\citep{DellaValle+06,Fynbo+06b,Gal-Yam+06} and require a paradigm shift in our understanding 
of the formation mechanism of GRBs or, at the very least, require the introduction of a new class of GRBs.
\item The proposed host galaxy of GRB 060614 is not the true host galaxy but, 
instead, represents the chance coincidence of a low-redshift galaxy intersecting 
the line of sight toward what is actually a moderate-redshift GRB (as suggested
by \citealt{SX06}).  
The SN, therefore, did occur but is undetectable given the redshift of the burst.
\end{enumerate}

In some ways, the fourth possibility is the most appealing, since it 
requires no new formation channel for long-duration GRBs, and is consistent
with some features of the GRB itself \citep{SX06}.
However, it requires a spatial coincidence between a foreground galaxy
and the GRB.  Here we analyze a deep optical image of the field of GRB 060614 to study
the statistics of such a coincidence empirically.  In agreement with \cite{SX06}, 
we find that a chance superposition is not implausible.  
While statistical arguments can never
rigorously exclude or require such a coincidence, the drastic nature of
the alternatives suggests that the possibility of a chance
superposition should be seriously considered.
  
\section{Data}
Our data were obtained using the ANDICAM instrument mounted on the 1.3 m telescope 
at Cerro Tololo Inter-American Observatory\footnote{http://www.astronomy.ohio-state.edu/ANDICAM}. 
This telescope is operated as part of the Small and Moderate Aperture Research Telescope System (SMARTS) 
consortium\footnote{http://www.astro.yale.edu/smarts.}.  
Imaging was obtained approximately every other night between 2006 June 15 - July 25 UT, 
which is between 0.65 and 40.73 days post-burst.  
A number of observations were affected by clouds and are not included in this analysis.  
In total, 13 useful nights worth of data were collected (see Table 1).    

Each night's data set consisted of six individual 360 s I-band observations and $30\times60$ s J-band
observations.  
The data were reduced in the same way as in \cite{Cobb+06} 
and combined to produce a single 36 minute I-band exposure and a 30 minute J-band exposure per night.
I-band observations taken 0.65 and 1.66 days post-burst both contain bright afterglow light 
and the remaining 11 images show no 
indication of brightening due to a SN \citep[e.g.][]{CB06} down
to 22 mag in I and 20 mag in J (see Table 1 and Figure 1).  This result is in agreement with the more stringent limits 
imposed by other works \citep{DellaValle+06,Fynbo+06b,Gal-Yam+06}.

The 11 I-band images taken from 2.66 to 40.73 days post-burst were combined to produce a single image of a 
$5\farcm4 \times 5\farcm2$ field approximately centered on GRB 060614.  
The I-band photometry is calibrated to a number of 
secondary standard stars in the field of GRB 060614.  The magnitude of these secondary standards 
was derived using Landolt standard star observations \citep{Landolt92} taken on 7 of the 
photometric nights when GRB 060614 was observed.   
The image reaches a $3\sigma$ limiting magnitude of $I\approx23.3$.   
The proposed host galaxy has an observed magnitude of $I = 22.08 \pm 0.09$.

All objects in the field were cataloged using SExtractor \citep{BA96}.  
348 objects were detected using a 2$\sigma$ detection threshold.
Using SExtractor's neural network star/galaxy classifier, 85 objects with CLASS\_STAR$>0.8$ 
were considered stars and not included in the following analysis.   
A section of our master image is shown in Figure 2 with SExtracted galaxies shown as 
green and magenta ellipses.  The proposed host galaxy of GRB 060614 is highlighted in red.     

\section{Probability of a Chance Superposition}
In our 28 square arcminute field of view, the position of the optical afterglow of GRB 060614 
can be determined to within sub-pixel accuracy.  
We calculate the probability by dividing the number of pixels covered by galaxies by
the total number of pixels.
The total pixel area of our image that is covered by galaxies is found by summing the
number of pixels contained in each individual SExtracted galaxy.  The area of each
galaxy can be computed in two ways.  First, the galaxy area can be taken as the area 
of the SExtracted ellipse ($\pi\times3*$A\_IMAGE$\times3*$B\_IMAGE),
where the multiplier 3 is chosen so that the ellipse is visually coincident with the extent of the galaxy (see Figure 2).  
Alternatively, the area can be given by the isophotal SExtractor parameter ISOAREAF\_IMAGE.  From the sum of the ellipse areas,
the number of pixels contained in galaxies divided by the total number of pixels in the image produces a probability of 3.5\% 
that the optical afterglow would land on a pixel contained in any galaxy in the field.  The isophotal area 
measurement produces somewhat smaller area values for each galaxy, reducing the probability to 2.4\%.  

Of course, some of the galaxies in this image are dimmer than the proposed host galaxy of GRB 060614, 
and we are concerned here specifically with the chance of a superposition with galaxies similar to
or brighter than the proposed host.
All galaxies that are detected at the $4.8\sigma$
level are considered ``bright" galaxies (these galaxies are indicated in Figure 2 by the green ellipses).
The value of $4.8\sigma$ is chosen because it is the highest
threshold for which the proposed host of GRB 060614 is still SExtracted.  Of the 263 galaxies
initially detected, 95 of those objects are still detected at a significance of greater
than $4.8\sigma$. The isophotal area contained within only these bright galaxies
represents 1.9\% of the total pixels in the image.  The number of galaxies
in this sample is reduced to 36\% of the original, but this corresponds
to only a 21\% drop in probability because each bright galaxy covers
significantly more area than any given dim galaxy.

To more fully explore the possibility that GRB 060614 is a chance superposition with a foreground
galaxy, it would be helpful to know the redshift of all the sources in the field.  In the
absence of this data, we can only assume for this field a typical galaxy redshift distribution.  
In a VVDS-CDFS galaxy sample limited to $I<22.1$ \citep{LeFevre+04}, 99\%
of galaxies have a redshift of $z\leq1.5$, which is at the low end of the redshift
range for GRB 060614 suggested by \cite{SX06}.  This indicates that if the burst was 
randomly associated with any of these bright galaxies, then that bright galaxy would almost certainly be at a redshift 
lower than that proposed for the burst.  However, it was only the lack of an observed SN
that called attention to this burst. If the proposed host galaxy of GRB 060614 had been at a redshift beyond that for which a SN could
reasonably have been detected, then the burst would not have been singled out as either curious
or groundbreaking given the burst's otherwise unexceptional characteristics.  

There exists spectroscopic
and/or photometric evidence for a SN component in most other bursts with $z\lesssim0.7$ \citep{ZKH04,Ferrero+06}.
However, at redshifts approaching $z\sim0.7$, the degeneracy between distance
and reddening would allow for the reasonable assumption that the SN
was not observed due to moderate extinction by host-galaxy dust.
To some extent, reddening can be constrained by analysis
of the X-ray to optical SED of the GRB afterglow.  In the case of GRB 060614, this yields
a low reddening value of only a few tenths of a magnitude \citep{DellaValle+06}.  
This assumes, however, that the GRB afterglow experiences the same reddening as the associated SN, 
which may not be the case given the collimated versus (near-)isotropic configuration of the GRB and SN emission, respectively.
The exact redshift at which a SN non-detection would have elicited serious proposals of a new GRB mechanism
is difficult to quantify, but we choose a value of $z=0.4$. At this redshift, the non-detection of a SN excess above
the brightness of a $I\sim22$ host galaxy requires that any SN be at least a magnitude fainter
at peak than the typical GRB-SNe value of $I\sim-19$, assuming a low reddening value measured from SED analysis.
Given that a variation of only a magnitude is not surprising for Type Ibc SNe in general, this limit
is still reasonable because HST observations may be expected to place more stringent
limits than ground-based observations by resolving the optical afterglow/SN location within
the galaxy.  In such a case, the SN would be observed as a deviation in the
power-law decay of the optical afterglow.  For example, a SN at $z=0.7$ could have 
been detected above the afterglow in the HST observations taken by \cite{Gal-Yam+06}.

In our flux-limited sample of VVDS-CDFS galaxies, 65\% have a redshift of $z\leq0.7$, and 25\% have $z\leq0.4$.  If projected
area is similar, or at least random, for all galaxies, then the area covered by those galaxies is 
2/3 or 1/4 of the total covered area, respectively.  The specific probability of a chance superposition with a galaxy that
has a redshift low enough for a null-SN detection to be of interest is then 1.2\% (for $z\leq0.7$) or 0.5\% (for $z\leq0.4$).
This does not take into account, however, the true angular size of the individual galaxies.  If the lowest redshift galaxies
trend toward larger projected surface areas than higher-$z$ galaxies, then the drop in the chance superposition
probability caused by this redshift cut will be somewhat reduced.

Thus, depending on one's assumptions of the redshift limit out to which the lack of an
apparent SN would be flagged as important, a chance superposition for this particular GRB can be ruled 
out with 98.1 - 99.5\% certainty.

\section{Discussion}
Although the probability for a chance superposition for any given GRB is small,
we must also consider that a large number of GRB fields have now been observed
in this manner.  If we assume that the galaxy density in the field of GRB 060614 is not unusual
(as appears to be the case --- Cobb et al. in prep), then given the $\approx 180$ bursts observed
by \textit{Swift}, a chance overlap of a galaxy brighter than $I\sim22$ and the
precise position of a GRB would be expected for $\approx 1 - 4$ GRBs.
The probability that no chance overlaps would be observed in the ensemble is
quite small, at 3.2\% to 41\%, assuming a probability of 1.9\% to 0.5\% for a 
chance superposition in any given image.  Most such superpositions would not be noticed, however, because the
typical foreground galaxy would be at a redshift comparable to that of the GRB.  In contrast,
GRB 060614 has generated great interest because the galaxy implies that
the GRB is at an unusually low redshift.

The strongest host galaxy claims can only be made for GRBs with 
detected optical afterglows.  Of the $\sim80$ optical afterglows detected following
\textit{Swift} bursts, the detection of no chance overlaps would not be unexpected
at a probability of 22\% to 67\%.  Note, however, that the position of GRBs without 
optical afterglows are determined from \textit{Swift} XRT
observations of X-ray afterglows, which have a spatial precision of $\sim4"$.
Taking this larger positional uncertainty into account would significantly increase the probability
of chance superpositions.

One additional feature of GRB 060614 is that it appears to be at the periphery of
the galaxy \citep{Gal-Yam+06}.  This is not expected for
standard long-duration GRB formation mechanisms, since the star formation rate at the outskirts of galaxies
is expected to be low.  Indeed \cite{Gal-Yam+06} invoke the location of the GRB within the
galaxy as a first clue to the nature of a new GRB formation paradigm.  The chance 
superposition hypothesis is consistent with an off-center location of the GRB.
With deeper observations it may be possible to confirm or refute the presence of a background galaxy,
although we note that there is a $\sim1/4$ chance that the center of the true host
galaxy will lie directly behind that of the $z=0.125$ galaxy \citep{Bloom+02}.  

We note that any superposition with a foreground galaxy requires that the
source be lensed.  However, the low luminosity and hence inferred low mass for this
particular galaxy suggests that the lensing induced magnification is less than 5\%,
which is inconsequential.  This lensing estimate is obtained using standard
assumptions and modeling the galaxy as a singular isothermal sphere.

It has been suggested that GRB 060505 is also a low-redshift GRB with
no SN \citep{Fynbo+06b}, although the observational data are more
limited both in the gamma-rays \citep{SX06} and for the afterglow,
for which no optical spectrum has been reported.  Clearly the existence
of a second ``060614-like" object would significantly diminish
the probability of chance superpositions.  But this argument
requires that the two objects be members of the \textit{same}
putative new category, and it is not clear that this could
be the case.  In particular, there are significant differences
in the GRB characteristics of the two sources (GRB 060505 was
much shorter--4 s versus 120 s--and weaker than GRB 060614; \citealt{Palmer+06,Parsons+06})
and GRB 060505 appears to be located in a star forming region of a galactic
spiral arm, whereas GRB 060614 seems to be outside the main
region of star formation.  If, as seems likely, the inferred
low redshifts of these two objects require different
formation mechanisms from each other, then an explanation of either
or both of them as a chance superposition remains plausible.

We emphasize that an analysis of this kind is by nature not definitive. 
From this argument alone, it is not possible to claim that GRB 060614 \textit{is} a chance superposition
between a low-$z$ galaxy and a moderate-$z$ GRB; we only claim that there
is a possibility of such an occurrence.
The fact that galaxies as bright as the purported host of GRB 060614 cover 
several percent of the field, and that well over a hundred GRBs have now been observed
by \textit{Swift}, suggests that a chance superposition cannot be excluded.  Indeed, over
the ensemble of \textit{Swift} bursts, the probability that such a coincidence would be
observed in at least one case is high.  
The unusual combination of a low-redshift putative host and the lack of an associated
supernova make GRB 060614 an intriguing candidate for a chance superposition.
Thus the conclusion that GRB 060614 requires
a ``new paradigm" for GRB formation should be approached with caution.

\acknowledgments
We thank SMARTS observers D. Gonzalez and J. Espinoza for their dedication to observing this source
and S. Tourtellotte for assistance with
optical data reduction. This work is supported by NSF
Graduate Fellowship DGE0202738 to BEC and NSF/AST grant 0407063 and \textit{Swift}
grant NNG05GM63G to CDB.

\begin{deluxetable}{rrll}
\tablecolumns{3}
\tablewidth{0pc}
\tablecaption{SMARTS observations of GRB 060614 afterglow + purported host galaxy}
\tablehead{\colhead{Observation UT date (2006)} & \colhead{Days after GRB\tablenotemark{a}} & \colhead{I mag\tablenotemark{b}} & \colhead{J mag\tablenotemark{b,c}}
}
\startdata
    June 15.1780 &       0.65 & $19.00\pm0.04$ 	& $18.39\pm0.11$	\\
    June 16.1909 &       1.66 & $20.62\pm0.10$ 	& $19.43\pm0.19$	\\
    June 17.1865 &       2.66 & $21.41\pm0.15$ 	& $>19.1$		\\
    June 19.3164 &       4.79 & $>21.7$ 	& $>19.0$		\\
    June 21.3072 &       6.78 & $>21.9$ 	& $>19.1$		\\
    June 25.4049 &      10.87 & $>21.8$ 	& $>19.1$		\\
    June 27.3962 &      12.87 & $22.06\pm0.14$ 	& $>19.8$		\\
    June 29.3504 &      14.82 & $22.04\pm0.15$ 	& $>18.9$		\\
    July 01.3760 &      16.85 & $>21.8$ 	& $>19.2$		\\
    July 05.2827 &      20.75 & $>22.0$ 	& $>18.9$		\\
    July 18.2019 &      33.67 & $22.13\pm0.13$ 	& $>19.9$		\\
    July 20.1319 &      35.60 & $22.31\pm0.21$ 	& $>19.6$		\\
    July 25.2559 &      40.73 & $22.09\pm0.14$ 	& $>20.0$		\\
\enddata
\tablenotetext{a}{Mid-exposure time of images in days after the burst trigger, which occurred at 2006 June 14, 12:43:48 UT.}
\tablenotetext{b}{These values have not been corrected for Galactic extinction. $3\sigma$ limiting magnitudes are reported for observations
in which the galaxy is not significantly detected.}
\tablenotetext{c}{J-band photometric calibration was performed using two 2MASS stars \citep{Skrutskie+06} in the field.}
\end{deluxetable}

\begin{figure}
\includegraphics[width=1\textwidth]{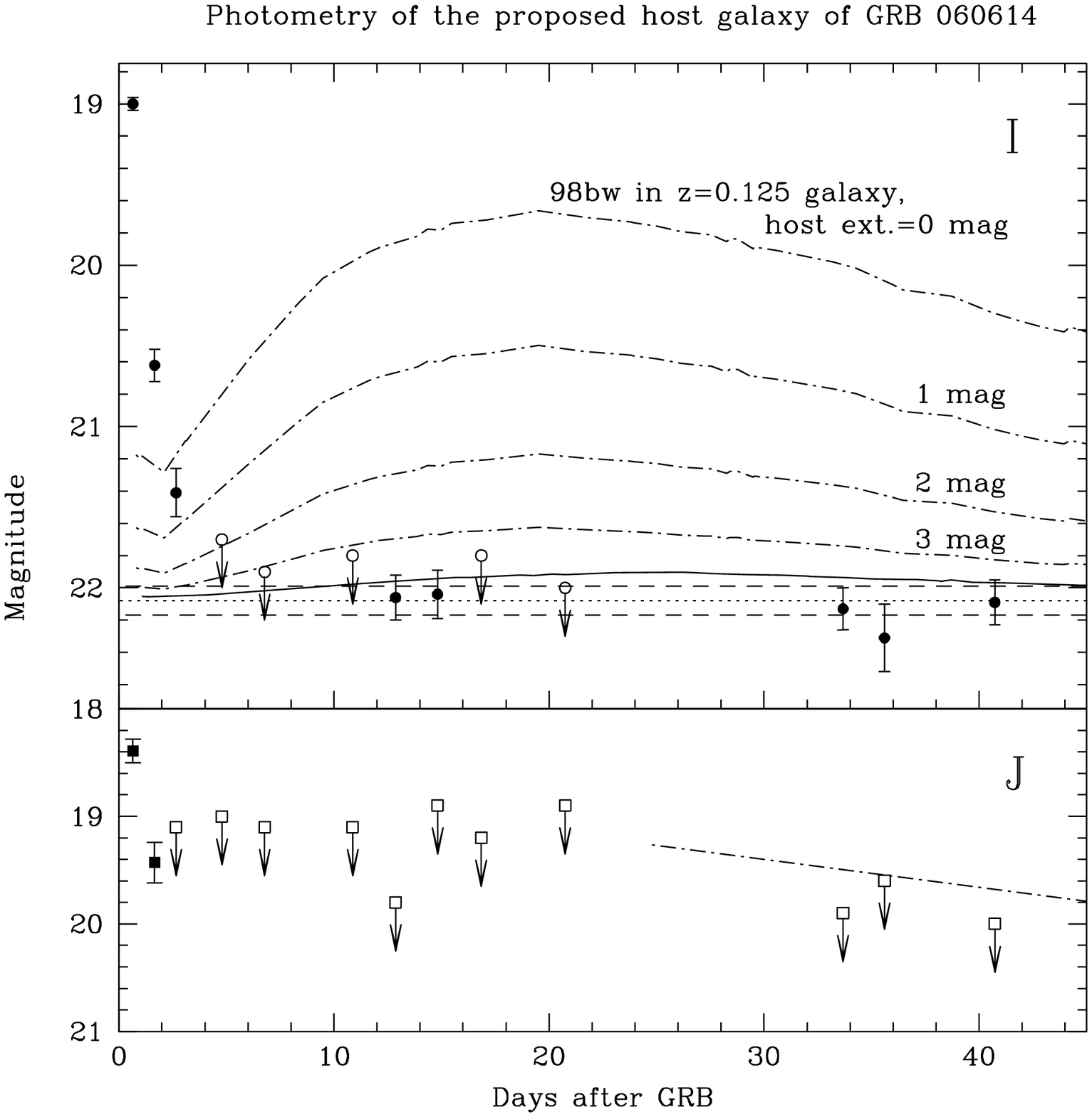}
\caption{Photometry of the afterglow and proposed host galaxy of GRB 060614.
Filled symbols represent detections; open symbols indicate $3\sigma$ magnitude limits
for observations in which the galaxy was not significantly detected.  
\textit{Top:} I-band photometry, with the brightness of the galaxy
determined from the master combined image ($I=22.08\pm0.09$) indicated by the dotted line with $1\sigma$
error (dashed lines).  The dot-dashed lines indicate 
the expected brightness of the galaxy had a SN similar to SN 1998bw \citep{Galama+98}
occurred in the galaxy, assuming 0 to 3 magnitudes of host-galaxy extinction.  
Even with 3 magnitudes of extinction, at $z=0.125$ such a SN would have 
easily been observed. The solid line indicates a $z=0.5$ SN 1998bw superimposed on the galaxy;
a SN with any higher redshift would be difficult to detect in our data.  Note
that other more stringent limits are available in the literature \citep{DellaValle+06,Fynbo+06b,Gal-Yam+06}.
\textit{Bottom:} J-band photometry; in the absence of the GRB afterglow, the galaxy remains
undetected.  A 1998bw-like SN would be expected to roughly follow the dot-dashed line, 
assuming the galaxy has a J-band magnitude of $\sim21$ and no host-galaxy extinction.  The few available SN 1998bw
J-band observations are taken from \cite{Patat+01}. 
}   
\end{figure}

\begin{figure}
\includegraphics[width=1\textwidth]{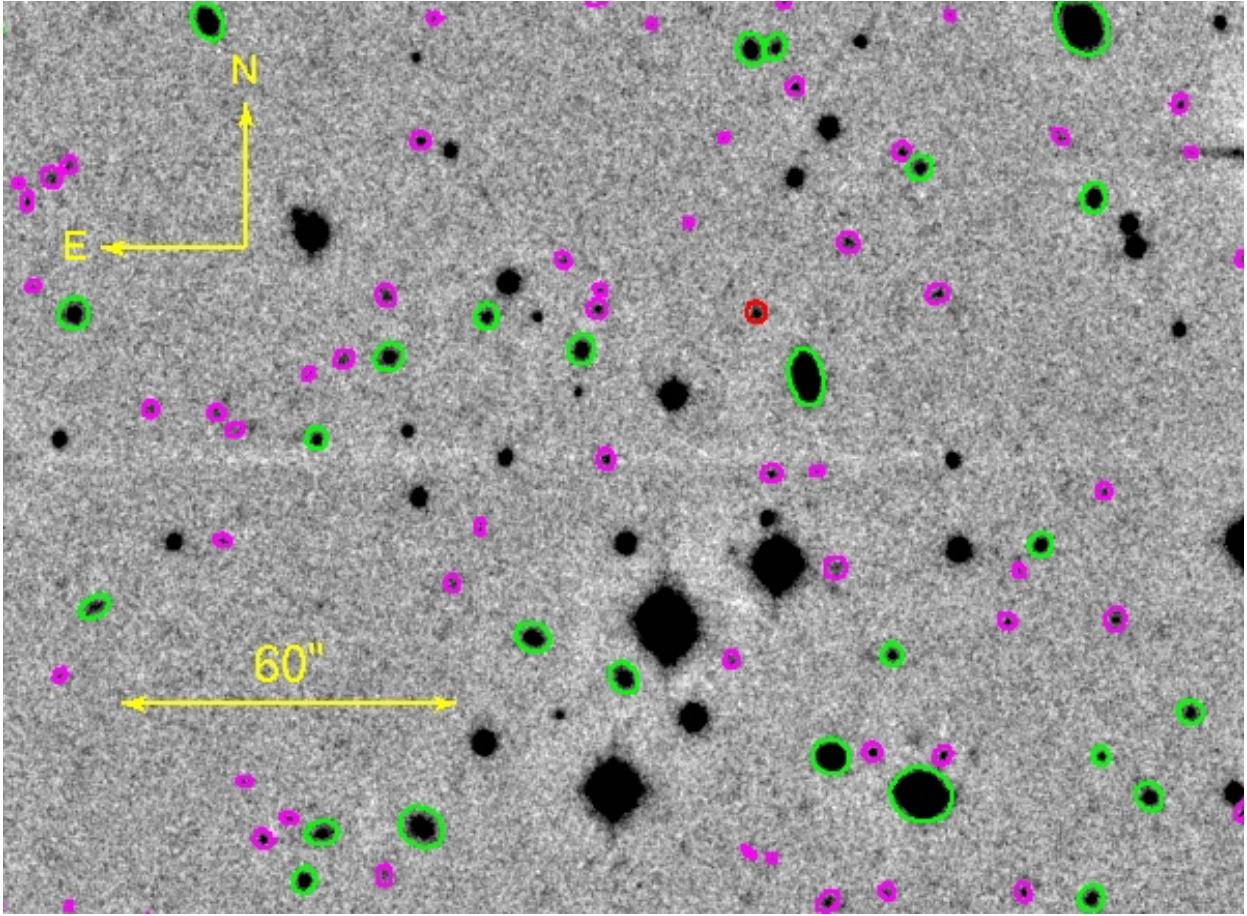}
\caption{Section of the master image of GRB 060614.  SExtracted galaxies are indicated in green
and magenta.  Each ellipse has axes 3*A\_IMAGE and 3*B\_IMAGE and position angle THETA\_IMAGE.
The proposed host galaxy of GRB 060614 is indicated in red.  Bright galaxies are denoted 
by green ellipses; these are galaxies detected at the same or higher significance 
than the galaxy at the position of the optical afterglow of GRB 060614.
}
\end{figure}
\end{document}